\documentclass[preprint, superscriptaddress,showpacs,preprintnumbers,amsmath,amssymb]{revtex4}
\usepackage{graphicx}
\usepackage{amsmath}%\usepackage{slashed}

\begin{document}

\thispagestyle{empty}

\title{Quantum electrodynamic approach to the conductivity of gapped graphene}

\author{
G.~L.~Klimchitskaya}
\affiliation{Central Astronomical Observatory at Pulkovo of the Russian Academy of Sciences,
Saint Petersburg,
196140, Russia}
\affiliation{Institute of Physics, Nanotechnology and
Telecommunications, Peter the Great Saint Petersburg
Polytechnic University, Saint Petersburg, 195251, Russia}

\author{
V.~M.~Mostepanenko}
\affiliation{Central Astronomical Observatory at Pulkovo of the Russian Academy of Sciences,
Saint Petersburg,
196140, Russia}
\affiliation{Institute of Physics, Nanotechnology and
Telecommunications, Peter the Great Saint Petersburg
Polytechnic University, Saint Petersburg, 195251, Russia}
\affiliation{Kazan Federal University, Kazan, 420008, Russia}
\begin{abstract}
The electrical conductivity of graphene with a nonzero mass-gap
parameter is investigated starting from the first principles of
quantum electrodynamics in (2+1)-dimensional space-time at any
temperature. The formalism of the polarization tensor defined
over the entire plane of complex frequency is used. At zero
temperature we reproduce the results for both real and imaginary
parts of the conductivity, obtained previously in the local
approximation, and generalize them taking into account the
effects of nonlocality. At nonzero temperature the exact
analytic expressions for real and imaginary parts of the
longitudinal and transverse conductivities of gapped graphene
are derived, as well as their local limits and approximate
expressions in several asymptotic regimes. Specifically, a
simple local result for the real part of conductivity of
gapped graphene valid at any temperature is obtained. According
to our results, the real part of the conductivity is not equal
to zero for frequencies exceeding the width of the gap and
goes to the universal conductivity with increasing frequency.
The imaginary part of conductivity of gapped graphene varies
from infinity at zero frequency to minus infinity at the
frequency defined by the gap parameter and then goes to zero
with further increase of frequency. The analytic expressions
are accompanied by the results of numerical computations.
Possible future generalization of the used formalism is
discussed.
\end{abstract}

\maketitle

\section{Introduction}
During the last few years, a two-dimensional hexagonal lattice of
carbon atoms known as graphene has attracted widespread attention
owing to its unusual
physical properties and promising applications \cite{1}.
Unlike conventional condensed matter systems, the quasiparticles
in graphene are massless or very light particles which at low
energies obey the relativistic Dirac equation, where the speed of light
$c$ is replaced with the Fermi velocity $v_F\approx c/300$
(see the monograph \cite{1} and review papers \cite{2,3,4}).
This makes the electronic properties of graphene highly
nontrivial, and, specifically, results in the existence \cite{5}
of a universal (up to small nonlocal corrections)
frequency-independent
conductivity $\sigma_0$ in the limit of zero temperature.

The electrical conductivity of graphene was investigated in many
theoretical papers using the current-current correlation functions
 in the random phase approximation, the two-dimensional Drude
model, the quasiclassical approach, the Kubo formula, and
Boltzmann's transport theory (see, e.g.,
Refs.~\cite{6,7,8,9,10,11,12,13,14,15,16,17,18,19,20,21,22,23,24,25,26,27,28,29,30}).
At first, a few distinct values for the universal conductivity
$\sigma_0$ have been obtained by different authors (see
Refs.~\cite{4,9,10,21} for a review) depending on the order
of limiting transitions in the used theoretical approach.
At a later time, a consensus was achieved on the value
$\sigma_0=e^2/(4\hbar)$ \cite{5,6,13,18,21,27,30}.
However, up to now there is no complete agreement between the
results of the different approximate and phenomenological approaches
to the conductivity of graphene at nonzero temperature,
in the presence of nonzero mass-gap parameter and chemical
potential. In doing so, the experimental results
\cite{31,32,33,34,35}
are not of sufficient precision for a clear discrimination between
the various theoretical predictions.

Quantum electrodynamics at nonzero temperature provides a regular
way for describing the response of a physical system to the
electromagnetic field by means of polarization tensor.
For graphene with an arbitrary mass-gap parameter $\Delta$ an
explicit expression for the polarization tensor in
(2+1)-dimensional
space-time was found in Ref.~\cite{36} at zero temperature.
The temperature correction to the polarization tensor of graphene
was derived in Ref.~\cite{37} at the pure imaginary Matsubara
frequencies. The results of Refs.~\cite{36,37} have been used to
calculate the Casimir and Casimir-Polder forces
\cite{37a,37b} in many real
physical systems incorporating graphene sheets
\cite{38,39,40,41,42,43,44,45} (previously such calculations
were performed by means of the density-density correlation
functions of graphene, the spatially nonlocal dielectric
permittivities, phenomenological Drude model, etc.
\cite{46,47,48,49,50,51}). The polarization tensor of graphene
was also used \cite{41,52} to compare with theory the experimental
data on measuring the Casimir interaction in graphene systems
\cite{53}, and very good agreement was obtained.

Another representation for the polarization tensor of graphene
valid over the entire plane of complex frequency was derived  in
Ref.~\cite{54}. It was applied to investigate the origin of large
thermal effect in the Casimir interaction between two graphene
sheets \cite{55,55a} and to calculate the reflectivity properties of
gapless graphene \cite{54}.
In Ref.~\cite{57} the same polarization tensor allowed
calculation of the reflectivity properties of graphene-coated
substrates.

The derivation of the polarization tensor valid along the real
frequency axis made it possible to investigate the electrical
conductivity of graphene on the basis of first principles of
quantum electrodynamics at nonzero temperature.
Note that in Ref.~\cite{54} explicit expressions for the
polarization tensor at real frequencies were obtained only for
a pure (pristine) graphene having a zero mass-gap parameter,
whereas for a gapped graphene the respective expressions were
presented in an implicit form. Because of this, as the starting point,
the conductivity of pure graphene was investigated \cite{58}.
In doing so, a few results
obtained previously using different theoretical approaches
have been reproduced, and several novel analytic asymptotic
expressions and numerical results (especially for the imaginary
part of conductivity) have been obtained.

In this paper we apply the formalism of the polarization tensor
to systematically investigate the electrical conductivity of
gapped graphene on the solid foundation of quantum
electrodynamics.
For this purpose, an explicit form of the results of
Ref.~\cite{54} related to the case of nonzero mass-gap parameter
is employed, which was found in Ref.~\cite{56}.
It is known that quasiparticles in graphene may acquire some
small mass $m$ under the influence of impurities,
electron-electron interactions, and in the presence of
substrates \cite{2,59,60,61}. The conductivity of gapped
graphene was studied in Ref.~\cite{20} using the two-band model
and in Ref.~\cite{29} using the static polarization function.
At zero temperature some analytic expressions for the
conductivity of gapped graphene were obtained in
Refs.~\cite{16,17},
whereas Ref.~\cite{27} contains the final results for both real
and imaginary parts of conductivity in the local approximation.
We reproduce the zero-temperature analytic results of
Ref.~\cite{27} within the formalism of the polarization tensor
and generalize them taking into account the effects of
nonlocality. Specifically, it is shown that at zero temperature
the real part of the conductivity of gapped graphene is equal
to zero in the frequency region from zero to the
characteristic frequency,
 which is slightly larger than $\Delta/\hbar$ due to nonlocal
effects. With increasing frequency the real part of conductivity
decreases monotonously from $2\sigma_0$ to $\sigma_0$.

According to our results, nonzero temperature has a dramatic
impact on the conductivity of gapped graphene. Thus, the real
part of conductivity for frequencies larger than the characteristic
frequency may be either  decreasing or increasing
function depending on the relationship between the mass-gap
parameter and the temperature. The imaginary part of the
conductivity of gapped graphene varies from plus to minus
infinity when frequency increases from zero to the characteristic
frequency (at zero temperature it varies from zero to minus
infinity in the same frequency region). We obtain the analytic
asymptotic expressions and perform numerical computations  for
both real and imaginary parts of the conductivity of gapped
graphene at nonzero temperature.

The paper is organized as follows. In Sec.~II, we present explicit
expressions for the polarization tensor of gapped graphene along
the real frequency axis. In Sec.~III, both real and imaginary
parts of the conductivity of gapped graphene are derived at zero
temperature. Sections IV and V investigate the impact of nonzero
temperature on real and imaginary parts of the conductivity of
gapped graphene, respectively. Our conclusions and discussion
are contained in Sec.~VI.

\section{Polarization tensor of gapped graphene}

Here, we present analytic results for the polarization tensor of
gapped graphene along the real frequency axis derived in
Refs.~\cite{54,56}. We also demonstrate how real and imaginary
parts of the conductivity of graphene are expressed via imaginary and
real parts of the polarization tensor, respectively.
Both the longitudinal (along the graphene surface) and transverse
(normal to the graphene surface) conductivities are considered.

The description of graphene using the polarization tensor in the
one-loop approximation and the density-density correlation
functions in the random phase approximation are equivalent
 \cite{42}. The former method is, however, somewhat advantageous
because it relies on well understood formalism of quantum field
theory at nonzero temperature. This allowed complete derivation
of the polarization tensor of gapped graphene valid over the
entire plane of complex frequency \cite{54}. In Ref.~\cite{56}
a more convenient form of this tensor especially along the real
frequency axis has been obtained for a gapped graphene.

As usual, we notate the polarization tensor as
$\Pi_{\mu\nu}(\omega,k,T)$, where $\mu,\,\nu=0,\,1,\,2$,
$\omega$ is the frequency of an electromagnetic wave, $k$ is the
magnitude of the wave vector component parallel to graphene,
and $T$ is the temperature. We assume real photons on a
mass-shell,
so that the inequality $k\leq\omega/c$ is satisfied. It has been
shown \cite{36,37} that only the two components of the polarization
tensor of graphene are independent. It is conventional to use
$\Pi_{00}$ and $\Pi_{\rm tr}\equiv\Pi_{\mu}^{\,\mu}$ as the two
independent quantities. For our purposes, however, it is more
convenient to use the following combination
\begin{equation}
\Pi(\omega,k,T)\equiv k^2\Pi_{\rm tr}(\omega,k,T)+
\left(\frac{\omega^2}{c^2}-k^2\right)\Pi_{00}(\omega,k,T)
\label{eq1}
\end{equation}
\noindent
as the second quantity.

We present $\Pi_{00}$ and $\Pi$ as the sums of zero-temperature
contributions and temperature corrections
\begin{eqnarray}
&&
\Pi_{00}(\omega,k,T)=\Pi_{00}^{(0)}(\omega,k)+
\Delta_{T}\Pi_{00}(\omega,k,T),
\nonumber \\
&&
\Pi(\omega,k,T)=\Pi^{(0)}(\omega,k)+
\Delta_{T}\Pi(\omega,k,T).
\label{eq2}
\end{eqnarray}
\noindent
The thermal corrections $\Delta_{T}\Pi_{00}$ and $\Delta_{T}\Pi$
vanish in the limit of zero temperature.

The analytic expressions for the polarization tensor of gapped
graphene at $T=0$ were derived in Ref.~\cite{36}. They result in
\begin{eqnarray}
&&
\Pi_{00}^{(0)}(\omega,k)=-\frac{\alpha k^2c^2}{\omega^2\eta^2}
\Phi(\omega,k),
\nonumber \\
&&
\Pi^{(0)}(\omega,k)=\alpha k^2\Phi(\omega,k),
\label{eq3}
\end{eqnarray}
\noindent
where $\alpha=e^2/(\hbar c)$ is the fine structure constant,
the quantity $\eta$ is defined as
\begin{equation}
\eta\equiv\eta(\omega,k)=\sqrt{1-\frac{v_F^2k^2}{\omega^2}}
\label{eq4}
\end{equation}
\noindent
and the function $\Phi$ along the real frequency axis
is given by \cite{54}
\begin{equation}
\Phi(\omega,k)=\left\{
\begin{array}{ll}
\frac{2\Delta}{c}-\frac{2\hbar\omega\eta}{c}\left[1+
\left(\frac{\Delta}{\hbar\omega\eta}\right)^2\right]
{\rm arctanh}\frac{\hbar\omega\eta}{\Delta}, &
\hbar\omega<\frac{\Delta}{\eta},
\\[3mm]
\frac{2\Delta}{c}-\frac{2\hbar\omega\eta}{c}\left[1+
\left(\frac{\Delta}{\hbar\omega\eta}\right)^2\right]
\left({\rm arctanh}\frac{\Delta}{\hbar\omega\eta}+
i\frac{\pi}{2}\right), &
\hbar\omega\geq\frac{\Delta}{\eta}.
\end{array}
\right.
\label{eq5}
\end{equation}
\noindent
Here, the energy gap $\Delta=2mc^2$ and $m$ is the  mass of a
quasiparticle (according to some estimations \cite{36} $mc^2$
may achieve 0.1\,eV).

As is seen from Eqs.~(\ref{eq3})--(\ref{eq5}), under the
condition $\hbar\omega\eta<\Delta$ the quantities $\Pi_{00}^{(0)}$
and $\Pi^{(0)}$ are real. If the condition
$\hbar\omega\eta\geq\Delta$ is satisfied, they have both real and
imaginary parts.

We now continue by considering the thermal corrections in
Eq.~(\ref{eq2}).
The imaginary parts of both $\Delta_{T}\Pi_{00}$ and
$\Delta_{T}\Pi$
are also different from zero only under the condition
$\hbar\omega\eta\geq\Delta$ \cite{54,56}. The explicit forms
are the following \cite{54,56}:
\begin{eqnarray}
&&
{\rm Im}\Delta_T\Pi_{00}(\omega,k,T)=
\frac{8\alpha\hbar c^2}{v_F^2}\int_{u^{(-)}}^{u^{(+)}}\!\!\!
\frac{du}{e^{\beta u}+1}
\frac{(2cu-\omega)^2-v_F^2k^2}{\omega\eta
\sqrt{v_F^2k^2A(\omega,k)-(2cu-\omega)^2}},
\nonumber\\[-2mm]
&&\label{eq6} \\
&&
{\rm Im}\Delta_T\Pi(\omega,k,T)=
\frac{8\alpha\hbar\eta\omega}{v_F^2}\int_{u^{(-)}}^{u^{(+)}}\!\!\!
\frac{du}{e^{\beta u}+1}
\frac{(2cu-\omega)^2+v_F^2k^2[1-A(\omega,k)]}{
\sqrt{v_F^2k^2A(\omega,k)-(2cu-\omega)^2}},
\nonumber
\end{eqnarray}
\noindent
where
\begin{equation}
\beta\equiv\frac{\hbar c}{k_BT},\quad
A(\omega,k)\equiv 1-\left(\frac{\Delta}{\hbar\omega\eta}\right)^2
\label{eq7}
\end{equation}
\noindent
and the integration limits are
\begin{equation}
u^{(\pm)}\equiv u^{(\pm)}(\omega,k)=\frac{1}{2c}[\omega\pm
v_Fk\sqrt{A(\omega,k)}].
\label{eq8}
\end{equation}

The real parts of thermal corrections in Eq.~(\ref{eq2}) are
different
from zero at all frequencies. For the frequencies satisfying the
condition $\hbar\omega\eta<\Delta$ the results are \cite{56}
\begin{eqnarray}
&&
{\rm Re}\Delta_T\Pi_{00}(\omega,k,T)=
\frac{16\alpha\hbar c^2}{v_F^2}\int_{\Delta/(2\hbar c)}^{\infty}
\frac{du}{e^{\beta u}+1}
\left\{1-\frac{1}{2\omega\eta}\left[
\frac{(2cu+\omega)^2-v_F^2k^2}{
\sqrt{(2cu+\omega)^2-v_F^2k^2A(\omega,k)}}
\right.\right.
\nonumber \\
&&~~~~~~~
\left.\left.-
\frac{(2cu-\omega)^2-v_F^2k^2}{
\sqrt{(2cu-\omega)^2-v_F^2k^2A(\omega,k)}}
\right]\right\},
\nonumber\\[-2mm]
&&\label{eq9} \\
&&
{\rm Re}\Delta_T\Pi(\omega,k,T)=
\frac{16\alpha\hbar \omega^2}{v_F^2}\int_{\Delta/(2\hbar c)}^{\infty}
\frac{du}{e^{\beta u}+1}
\left\{1-\frac{\eta}{2\omega}\left[
\frac{(2cu+\omega)^2+v_F^2k^2[1-A(\omega,k)]}{
\sqrt{(2cu+\omega)^2-v_F^2k^2A(\omega,k)}}
\right.\right.
\nonumber \\
&&~~~~~~~
\left.\left.-
\frac{(2cu-\omega)^2+v_F^2k^2[1-A(\omega,k)]}{
\sqrt{(2cu-\omega)^2-v_F^2k^2A(\omega,k)}}
\right]\right\}.
\nonumber
\end{eqnarray}

Within the frequency region $\hbar\omega\eta\geq\Delta$
explicit representations for the quantities
${\rm Re}\Delta_T\Pi_{00}$ and ${\rm Re}\Delta_T\Pi$
are somewhat more complicated. Thus, according to Ref.~\cite{56},
the most convenient expression for ${\rm Re}\Delta_T\Pi_{00}$
in this region is the following:
\begin{equation}
{\rm Re}\Delta_T\Pi_{00}(\omega,k,T)=\frac{16\alpha\hbar
c^2}{v_F^2}(I_1+I_2+I_3),
\label{eq10}
\end{equation}
\noindent
where the integrals $I_{1,2,3}$ are defined as
\begin{eqnarray}
&&
I_1\equiv\int_{\Delta/(2\hbar c)}^{u^{(-)}}
\frac{du}{e^{\beta u}+1}
\left\{1-\frac{1}{2\omega\eta}\left[
\frac{(2cu+\omega)^2-v_F^2k^2}{
\sqrt{(2cu+\omega)^2-v_F^2k^2A(\omega,k)}}
\right.\right.
\nonumber \\
&&~~~~~~~
\left.\left.+
\frac{(2cu-\omega)^2-v_F^2k^2}{
\sqrt{(2cu-\omega)^2-v_F^2k^2A(\omega,k)}}
\right]\right\},
\nonumber\\[-2mm]
&&\label{eq11} \\
&&
I_2\equiv\int_{u^{(-)}}^{u^{(+)}}
\frac{du}{e^{\beta u}+1}
\left[1-\frac{1}{2\omega\eta}
\frac{(2cu+\omega)^2-v_F^2k^2}{
\sqrt{(2cu+\omega)^2-v_F^2k^2A(\omega,k)}}
\right],
\nonumber \\[1mm]
&&
I_3\equiv\int_{u^{(+)}}^{\infty}
\frac{du}{e^{\beta u}+1}
\left\{1-\frac{1}{2\omega\eta}\left[
\frac{(2cu+\omega)^2-v_F^2k^2}{
\sqrt{(2cu+\omega)^2-v_F^2k^2A(\omega,k)}}
\right.\right.
\nonumber \\
&&~~~~~~~
\left.\left.-
\frac{(2cu-\omega)^2-v_F^2k^2}{
\sqrt{(2cu-\omega)^2-v_F^2k^2A(\omega,k)}}
\right]\right\}.
\nonumber
\end{eqnarray}

In a similar way, the most convenient expression for
${\rm Re}\Delta_T\Pi$ in the frequency region
$\hbar\omega\eta\geq\Delta$
is given by \cite{56}
\begin{equation}
{\rm Re}\Delta_T\Pi(\omega,k,T)=
\frac{16\alpha\hbar\omega^2}{v_F^2}(J_1+J_2+J_3),
\label{eq12}
\end{equation}
\noindent
where the integrals $J_{1,2,3}$ are the following
\begin{eqnarray}
&&
J_1\equiv\int_{\Delta/(2\hbar c)}^{u^{(-)}}
\frac{du}{e^{\beta u}+1}
\left\{1-\frac{\eta}{2\omega}\left[
\frac{(2cu+\omega)^2+v_F^2k^2[1-A(\omega,k)]}{
\sqrt{(2cu+\omega)^2-v_F^2k^2A(\omega,k)}}
\right.\right.
\nonumber \\
&&~~~~~~~
\left.\left.+
\frac{(2cu-\omega)^2+v_F^2k^2[1-A(\omega,k)]}{
\sqrt{(2cu-\omega)^2-v_F^2k^2A(\omega,k)}}
\right]\right\},
\nonumber\\[-2mm]
&&\label{eq13} \\
&&
J_2\equiv\int_{u^{(-)}}^{u^{(+)}}
\frac{du}{e^{\beta u}+1}
\left[1-\frac{\eta}{2\omega}
\frac{(2cu+\omega)^2+v_F^2k^2[1-A(\omega,k)]}{
\sqrt{(2cu+\omega)^2-v_F^2k^2A(\omega,k)}}
\right],
\nonumber \\[1mm]
&&
J_3\equiv\int_{u^{(+)}}^{\infty}
\frac{du}{e^{\beta u}+1}
\left\{1-\frac{\eta}{2\omega}\left[
\frac{(2cu+\omega)^2+v_F^2k^2[1-A(\omega,k)]}{
\sqrt{(2cu+\omega)^2-v_F^2k^2A(\omega,k)}}
\right.\right.
\nonumber \\
&&~~~~~~~
\left.\left.-
\frac{(2cu-\omega)^2+v_F^2k^2[1-A(\omega,k)]}{
\sqrt{(2cu-\omega)^2-v_F^2k^2A(\omega,k)}}
\right]\right\}.
\nonumber
\end{eqnarray}

The above expressions (\ref{eq2})--(\ref{eq13}) provide an exact
representation for the polarization tensor of graphene at
arbitrary frequency, wave vector, temperature and mass-gap
parameter in the application region of the Dirac model.
This opens up possibility for a comprehensive investigation of
the conductivity of gapped graphene on the basis of first
principles of quantum electrodynamics. The point is that the
longitudinal and transverse conductivities of graphene are
directly connected with the respective density-density
correlation functions \cite{50,62}. The latter are expressed
via the components of the polarization tensor \cite{42}.
As a result, the longitudinal and transverse conductivities
of graphene take the form \cite{58,56}
\begin{eqnarray}
&&
\sigma_{\|}(\omega,k,T)=-i\frac{\omega}{4\pi\hbar k^2}
\Pi_{00}(\omega,k,T),
\label{eq14} \\
&&
\sigma_{\bot}(\omega,k,T)=i\frac{c^2}{4\pi\hbar k^2\omega}
\Pi(\omega,k,T).
\nonumber
\end{eqnarray}

In the next sections, using these equations, the conductivities
of gapped graphene are treated at both zero and nonzero
temperature.

\section{Conductivity of gapped graphene at zero temperature}

At $T=0$ the polarization tensor of gapped graphene is
characterized by the quantities
$\Pi_{00}^{(0)}$ and $\Pi^{(0)}$ which are presented in
Eqs.~(\ref{eq3})--(\ref{eq5}). We consider separately the real
and imaginary parts of the conductivity.

\subsection{Real part of conductivity}

According to Eq.~(\ref{eq14}), the real parts of graphene
conductivities
at zero temperature are given by
\begin{eqnarray}
&&
{\rm Re}\sigma_{\|}^{(0)}(\omega,k)=\frac{\omega}{4\pi\hbar k^2}
{\rm Im}\Pi_{00}^{(0)}(\omega,k),
\label{eq15} \\
&&
{\rm Re}\sigma_{\bot}^{(0)}(\omega,k)=-\frac{c^2}{4\pi\hbar k^2\omega}
{\rm Im}\Pi^{(0)}(\omega,k).
\nonumber
\end{eqnarray}

{}From Eqs.~(\ref{eq3})--(\ref{eq5}) one obtains
\begin{eqnarray}
&&
{\rm Im}\Pi_{00}^{(0)}(\omega,k)=
\frac{\alpha\pi\hbar ck^2}{\omega\eta}\left[1+
\left(\frac{\Delta}{\hbar\omega\eta}\right)^{\!\!2}\right]
\theta(\hbar\omega\eta-\Delta),
\nonumber \\[-2mm]
&&\label{eq16} \\
&&
{\rm Im}\Pi^{(0)}(\omega,k)=
-\frac{\alpha\pi\hbar k^2\omega\eta}{c}\left[1+
\left(\frac{\Delta}{\hbar\omega\eta}\right)^{\!\!2}\right]
\theta(\hbar\omega\eta-\Delta),
\nonumber
\end{eqnarray}
\noindent
where $\theta(z)$ is the step function.

Substituting Eq.~(\ref{eq16}) in Eq.~(\ref{eq15}) and taking
into account that $\alpha=e^2/(\hbar c)$, one arrives at
\begin{eqnarray}
&&
{\rm Re}\sigma_{\|}^{(0)}(\omega,k)=\frac{\sigma_0}{\eta}
\left[1+
\left(\frac{\Delta}{\hbar\omega\eta}\right)^{\!\!2}\right]
\theta(\hbar\omega\eta-\Delta),
\label{eq17} \\
&&
{\rm Re}\sigma_{\bot}^{(0)}(\omega,k)={\sigma_0}{\eta}
\left[1+
\left(\frac{\Delta}{\hbar\omega\eta}\right)^{\!\!2}\right]
\theta(\hbar\omega\eta-\Delta),
\nonumber
\end{eqnarray}
\noindent
where the universal conductivity of graphene is
\begin{equation}
\sigma_0=\frac{e^2}{4\hbar}.
\label{eq18}
\end{equation}

As is seen from Eq.~(\ref{eq17}), in the frequency region
satisfying the condition $\hbar\omega\eta<\Delta$ the real
parts of the conductivities of gapped graphene are equal to
zero. Note that the vanishing of the conductivity of gapped
graphene at low frequencies was noted in many papers using
various formalisms (see, for instance, Refs.~\cite{17,20,27}).

Taking into account that $k\leq\omega/c$, one obtains
\begin{equation}
\frac{v_F^2k^2}{\omega^2}\leq\frac{v_F^2}{c^2}
\approx 1.1\times 10^{-5}.
\label{eq19}
\end{equation}
\noindent
Using Eq.~(\ref{eq4}), we expand Eq.~(\ref{eq17}) in powers of
the small parameter (\ref{eq19}) and in the lowest order find
\begin{eqnarray}
&&
{\rm Re}\sigma_{\|}^{(0)}(\omega,k)={\sigma_0}\left\{
1+\left(\frac{\Delta}{\hbar\omega}\right)^{\!\!2}+
\frac{1}{2}\left(\frac{v_Fk}{\omega}\right)^{\!\!2}\left[
1+3\left(\frac{\Delta}{\hbar\omega}\right)^{\!\!2}
\right]\right\}
\theta(\hbar\omega\eta-\Delta),
\nonumber\\[-2mm]
&&\label{eq20} \\
&&
{\rm Re}\sigma_{\bot}^{(0)}(\omega,k)={\sigma_0}\left\{
1+\left(\frac{\Delta}{\hbar\omega}\right)^{\!\!2}-
\frac{1}{2}\left(\frac{v_Fk}{\omega}\right)^{\!\!2}\left[
1-\left(\frac{\Delta}{\hbar\omega}\right)^{\!\!2}
\right]\right\}
\theta(\hbar\omega\eta-\Delta).
\nonumber
\end{eqnarray}

As is seen from Eq.~(\ref{eq20}), ${\rm Re}\sigma_{\|}$ and
${\rm Re}\sigma_{\bot}$ differ only in the first perturbation
order in the small parameter (\ref{eq19})
and, thus, can be considered as equal for all practical
purposes. They are normalized to $\sigma_0$ and plotted in
Fig.~\ref{fg1} as functions of $\hbar\omega/\Delta$.
Note that in this and all the following figures computational
results for the longitudinal and transverse conductivities are
indistinguishable in the used scales. Because of this, they
are always presented by a single line.
In the region $\hbar\omega\geq\Delta$ the conductivities of
graphene are the  monotonous functions decreasing from $2\sigma_0$ to
$\sigma_0$ when the frequency increases from $\Delta/\hbar$ to
infinity.

In the local limit, where the results are independent on the
wave vector,  Eq.~(\ref{eq20}) leads to
\begin{equation}
{\rm Re}\sigma_{\|}^{(0)}(\omega,0)=
{\rm Re}\sigma_{\bot}^{(0)}(\omega,0)=
\sigma_0\frac{(\hbar\omega)^2+\Delta^2}{(\hbar\omega)^2}
\theta(\hbar\omega-\Delta).
\label{eq21}
\end{equation}
\noindent
Note that in our case the same results are obtained from
Eq.~(\ref{eq20}) taken at $k=0$.
They coincide with the results presented in Ref.~\cite{27}
obtained in the framework of another approach.

\subsection{Imaginary part of conductivity}

Using Eq.~(\ref{eq14}), for the imaginary part of conductivity
of graphene at zero temperature we obtain
\begin{eqnarray}
&&
{\rm Im}\sigma_{\|}^{(0)}(\omega,k)=-\frac{\omega}{4\pi\hbar k^2}
{\rm Re}\Pi_{00}^{(0)}(\omega,k),
\label{eq22} \\
&&
{\rm Im}\sigma_{\bot}^{(0)}(\omega,k)=\frac{c^2}{4\pi\hbar k^2\omega}
{\rm Re}\Pi^{(0)}(\omega,k).
\nonumber
\end{eqnarray}

For the real part of the polarization tensor from
Eqs.~(\ref{eq3})--(\ref{eq5}) one finds
\begin{eqnarray}
&&
{\rm Re}\Pi_{00}^{(0)}(\omega,k)=
-\frac{\alpha k^2c^2}{\omega^2\eta^2}{\rm Re}\Phi(\omega,k),
\nonumber \\
&&
{\rm Re}\Pi^{(0)}(\omega,k)=\alpha k^2{\rm Re}\Phi(\omega,k),
\label{eq23}
\end{eqnarray}
\noindent
where
\begin{equation}
{\rm
Re}\Phi(\omega,k)=\frac{2\Delta}{c}-\frac{2\hbar\omega\eta}{c}
\left[1+\left(\frac{\Delta}{\hbar\omega\eta}\right)^{\!\!2}\right]
\left\{
\begin{array}{ll}
{\rm arctanh}\frac{\hbar\omega\eta}{\Delta},&
\hbar\omega\eta<\Delta, \\
{\rm arctanh}\frac{\Delta}{\hbar\omega\eta},&
\hbar\omega\eta\geq\Delta.
\end{array}
\right.
\label{eq24}
\end{equation}

Using the identity \cite{63}
\begin{equation}
{\rm arctanh}z=\frac{1}{2}\ln\frac{1+z}{1-z},
\label{eq25}
\end{equation}
\noindent
Eq.~(\ref{eq24}) can be rewritten in the form
\begin{equation}
{\rm Re}\Phi(\omega,k)=
\frac{2\Delta}{c}-\frac{\hbar\omega\eta}{c}
\left[1+\left(\frac{\Delta}{\hbar\omega\eta}\right)^{\!\!2}\right]
\ln\left|\frac{\Delta+\hbar\omega\eta}{\Delta-\hbar\omega\eta}
\right|.
\label{eq26}
\end{equation}

Substituting Eqs.~(\ref{eq23}) and (\ref{eq26}) in
Eq.~(\ref{eq22}), one arrives at
\begin{eqnarray}
&&
{\rm Im}\sigma_{\|}^{(0)}(\omega,k)=\frac{\sigma_0}{\pi\eta}
\left[\frac{2\Delta}{\hbar\omega\eta}-
\frac{(\hbar\omega\eta)^2+\Delta^2}{(\hbar\omega\eta)^2}
\ln\left|\frac{\Delta+\hbar\omega\eta}{\Delta-\hbar\omega\eta}
\right|\right],
\nonumber\\[-2mm]
\label{eq27} \\
&&
{\rm Im}\sigma_{\bot}^{(0)}(\omega,k)=\frac{\sigma_0\eta}{\pi}
\left[\frac{2\Delta}{\hbar\omega\eta}-
\frac{(\hbar\omega\eta)^2+\Delta^2}{(\hbar\omega\eta)^2}
\ln\left|\frac{\Delta+\hbar\omega\eta}{\Delta-\hbar\omega\eta}
\right|\right].
\nonumber
\end{eqnarray}

{}From Eq.~(\ref{eq27}) it is seen that the nonlocal effects play
equally small role in the imaginary parts of conductivity of
graphene as in its real part. In the local limit we have
$\eta(\omega,0)=1$ and Eq.~(\ref{eq27}) reduces to
\begin{equation}
{\rm Im}\sigma_{\|}^{(0)}(\omega,0)={\rm Im}\sigma_{\bot}^{(0)}(\omega,0)=
\frac{\sigma_0}{\pi}
\left[\frac{2\Delta}{\hbar\omega}-
\frac{(\hbar\omega)^2+\Delta^2}{(\hbar\omega)^2}
\ln\left|\frac{\Delta+\hbar\omega}{\Delta-\hbar\omega}
\right|\right].
\label{eq28}
\end{equation}
\noindent
This coincides with the result presented in Ref.~\cite{27}.

\begin{figure}[b]
\vspace*{-6cm}
\centerline{\hspace*{.5cm}
\includegraphics{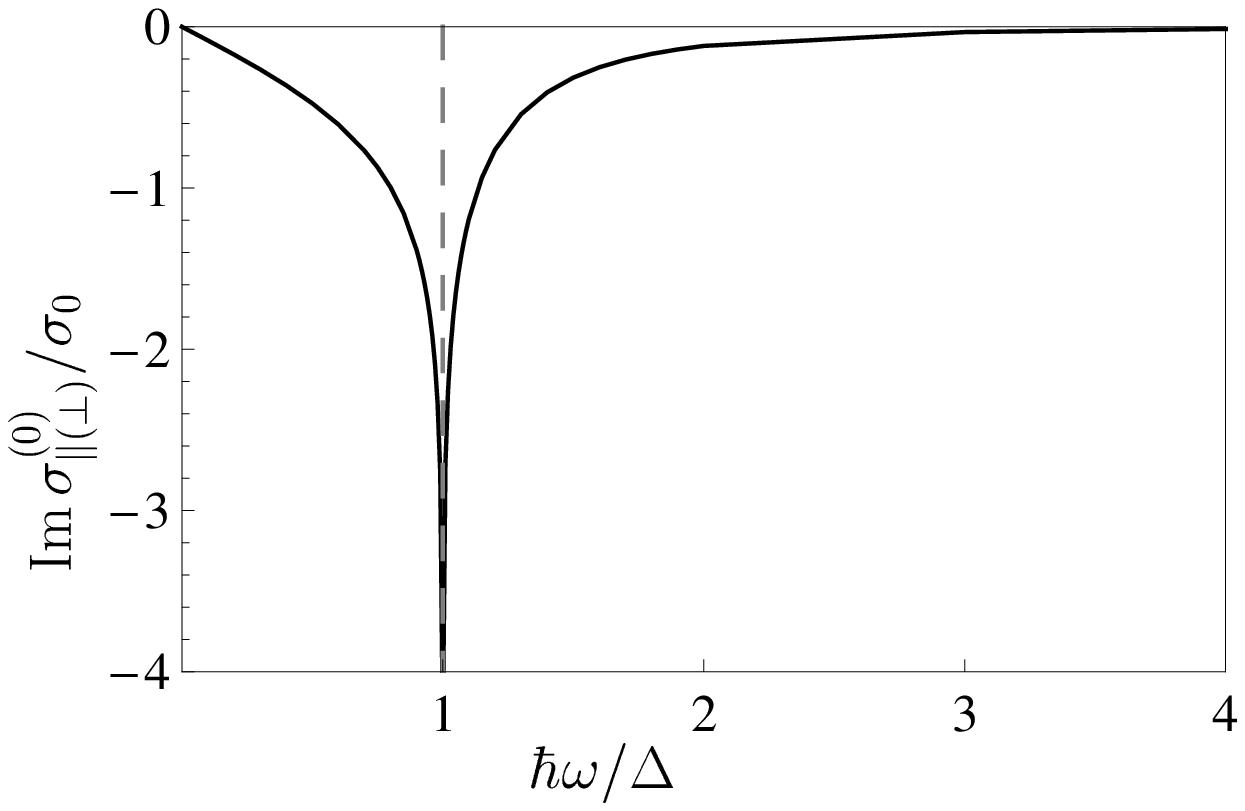}
}
\vspace*{-9cm}
\caption{\label{fg2}
The imaginary part of
the conductivity of gapped graphene
at zero temperature normalized to $\sigma_0$ is shown by the solid line
as a function of frequency divided by the width of the gap.
The dashed line illustrates the condition $\hbar\omega=\Delta$.
}
\end{figure}

In Fig.~\ref{fg2} we plot the imaginary
part of the conductivity of graphene
normalized to $\sigma_0$
at $T=0$ as a function of the ratio
of $\hbar\omega$ to the width of the gap $\Delta$.
As is seen in Fig.~\ref{fg2}, for any fixed $\Delta$ the
imaginary part of conductivity decreases monotonously from
zero to minus infinity when frequency increases from 0 to
$\Delta/\hbar$ and then increases to zero with further increase
of frequency.

\section{Real part of conductivity of gapped graphene at
nonzero temperature}

We start from the real parts of thermal corrections to the
conductivities of graphene. According to Eqs.~(\ref{eq2}) and
(\ref{eq14}), they are given by
\begin{eqnarray}
&&
{\rm Re}\Delta_T\sigma_{\|}(\omega,k,T)=\frac{\omega}{4\pi\hbar k^2}
{\rm Im}\Delta_T\Pi_{00}(\omega,k,T),
\label{eq29} \\
&&
{\rm Re}\Delta_T\sigma_{\bot}(\omega,k,T)=-\frac{c^2}{4\pi\hbar k^2\omega}
{\rm Im}\Delta_T\Pi(\omega,k,T).
\nonumber
\end{eqnarray}
\noindent
Recall that the quantities ${\rm Im}\Delta_T\Pi_{00}$ and
${\rm Im}\Delta_T\Pi$ (and, thus, ${\rm Re}\Delta_T\sigma_{\|}$
and ${\rm Re}\Delta_T\sigma_{\bot}$) are equal to zero under the
condition $\hbar\omega\eta<\Delta$.  Under the opposite
condition $\hbar\omega\eta\geq\Delta$ they are given by
Eq.~(\ref{eq6}).

It is convenient to introduce another integration variable in
Eq.~(\ref{eq6})
\begin{equation}
\tau=\frac{2cu-\omega}{v_Fk\sqrt{A(\omega,k)}},
\label{eq30}
\end{equation}
\noindent
where $A$ is defined in Eq.~(\ref{eq7}). Then Eq.~(\ref{eq6})
takes the form
\begin{eqnarray}
&&
{\rm Im}\Delta_T\Pi_{00}(\omega,k,T)=
-4\alpha\hbar\frac{ck^2}{\omega\eta}
\int_{-1}^{1}\frac{d\tau}{\exp\left(
\frac{\hbar\omega}{2k_BT}+\gamma\tau\right)+1}
\left[\frac{1-A(\omega,k)}{\sqrt{1-\tau^2}}+A(\omega,k)\sqrt{1-\tau^2}
\right],
\nonumber \\
&&\label{eq31} \\[-2mm]
&&
{\rm Im}\Delta_T\Pi(\omega,k,T)=
4\alpha\hbar\omega\eta\frac{k^2}{c}
\int_{-1}^{1}\frac{d\tau}{\exp\left(
\frac{\hbar\omega}{2k_BT}+\gamma\tau\right)+1}
\left[\frac{1}{\sqrt{1-\tau^2}}-A(\omega,k)\sqrt{1-\tau^2}
\right].
\nonumber
\end{eqnarray}
\noindent
Here, the parameter $\gamma$ is defined as
\begin{equation}
\gamma\equiv\gamma(\omega,k,T)=\frac{v_Fk}{\omega}\,
\frac{\hbar\omega}{2k_BT}\sqrt{A(\omega,k)}.
\label{eq32}
\end{equation}
\noindent
It is equal to zero in the local limit.

Expanding the exponent-containing fraction in the first formula
of Eq.~(\ref{eq31}) in a series and calculating the integrals
\cite{63} one obtains
\begin{eqnarray}
&&
{\rm Im}\Delta_T\Pi_{00}(\omega,k,T)=
-4\alpha\hbar\frac{ck^2}{\omega\eta}
\sum_{n=1}^{\infty}(-1)^{n-1}e^{-\frac{n\hbar\omega}{2k_BT}}
\int_{-1}^{1}d\tau e^{-n\gamma\tau}
\left[\frac{1-A(\omega,k)}{\sqrt{1-\tau^2}}+A(\omega,k)\sqrt{1-\tau^2}
\right]
\nonumber \\
&&\label{eq33} \\[-2mm]
&&~~
=
-4\alpha\hbar\frac{ck^2}{\omega\eta}
\sum_{n=1}^{\infty}(-1)^{n-1}e^{-\frac{n\hbar\omega}{2k_BT}}
 \left\{\left[1-A(\omega,k)\right]{ I}_0(n\gamma)+
\frac{A(\omega,k)}{n\gamma}{I}_1(n\gamma)\right\},
\nonumber
\end{eqnarray}
\noindent
where ${I}_k(z)$ are the Bessel functions of an imaginary
argument.

In a similar manner, the second formula of Eq.~(\ref{eq31}) can be
rewritten as
\begin{eqnarray}
&&
{\rm Im}\Delta_T\Pi(\omega,k,T)=
4\alpha\hbar\omega\eta\frac{k^2}{c}
\sum_{n=1}^{\infty}(-1)^{n-1}e^{-\frac{n\hbar\omega}{2k_BT}}
\int_{-1}^{1}d\tau e^{-n\gamma\tau}
\left[\frac{1}{\sqrt{1-\tau^2}}-A(\omega,k)\sqrt{1-\tau^2}
\right]
\nonumber \\
&&\label{eq34} \\[-2mm]
&&~~~~~~
=4\alpha\hbar\omega\eta\frac{k^2}{c}
\sum_{n=1}^{\infty}(-1)^{n-1}e^{-\frac{n\hbar\omega}{2k_BT}}
 \left[{ I}_0(n\gamma)-
\frac{A(\omega,k)}{n\gamma}{ I}_1(n\gamma)\right].
\nonumber
\end{eqnarray}

Substituting Eqs.~(\ref{eq33}) and (\ref{eq34}) with added
$\theta$-functions in Eq.~(\ref{eq29}),
we obtain the exact expressions for the real parts of
thermal corrections to longitudinal and transverse
conductivities of gapped graphene at nonzero temperature
\begin{eqnarray}
&&
{\rm Re}\Delta_T\sigma_{\|}(\omega,k,T)=-\frac{4\sigma_0}{\eta}
\theta(\hbar\omega\eta-\Delta)
\sum_{n=1}^{\infty}(-1)^{n-1}e^{-\frac{n\hbar\omega}{2k_BT}}
\left\{\left[1-A(\omega,k)\right]{ I}_0(n\gamma)+
\frac{A(\omega,k)}{n\gamma}{ I}_1(n\gamma)\right\},
\nonumber\\
&&
\label{eq35} \\[-2mm]
&&
{\rm Re}\Delta_T\sigma_{\bot}(\omega,k,T)=-{4\sigma_0}{\eta}
\theta(\hbar\omega\eta-\Delta)
\sum_{n=1}^{\infty}(-1)^{n-1}e^{-\frac{n\hbar\omega}{2k_BT}}
\left[{ I}_0(n\gamma)-
\frac{A(\omega,k)}{n\gamma}{ I}_1(n\gamma)\right].
\nonumber
\end{eqnarray}

To determinate the role of nonlocal effects we expand the
right-hand sides of both expressions in Eq.~(\ref{eq35}) up to
the first order in the small parameter (\ref{eq19}) and find
\begin{eqnarray}
&&
{\rm Re}\Delta_T\sigma_{\|}(\omega,k,T)=-\sigma_0
\theta(\hbar\omega\eta-\Delta)
\nonumber \\
&&
\times\left\{2\left[1+\left(\frac{\Delta}{\hbar\omega}\right)^{\!\!2}\right]
+\left(\frac{v_Fk}{\omega}\right)^{\!\!2}\!
\left[1+3\left(\frac{\Delta}{\hbar\omega}\right)^{\!\!2}\right]
\sum_{n=1}^{\infty}(-1)^{n-1}e^{-\frac{n\hbar\omega}{2k_BT}}\right.
\nonumber \\
&&~~~\left.
+\frac{1}{4}\left[1-\left(\frac{\Delta}{\hbar\omega}\right)^{\!\!2}\right]
\!\left[1+3\left(\frac{\Delta}{\hbar\omega}\right)^{\!\!2}\right]\!
\left(\frac{v_Fk}{\omega}\right)^{\!\!2}\!
\left(\frac{\hbar\omega}{2k_BT}\right)^{\!\!2}
\sum_{n=1}^{\infty}(-1)^{n-1}n^2e^{-\frac{n\hbar\omega}{2k_BT}}\right\},
\nonumber \\[-1mm]
&&\label{eq36} \\[-2mm]
&&
{\rm Re}\Delta_T\sigma_{\bot}(\omega,k,T)=-\sigma_0
\theta(\hbar\omega\eta-\Delta)
\nonumber \\
&&
\times\left\{2\left[1+\left(\frac{\Delta}{\hbar\omega}\right)^{\!\!2}\right]
-\frac{1}{2}\left(\frac{v_Fk}{\omega}\right)^{\!\!2}\!
\left[1-\left(\frac{\Delta}{\hbar\omega}\right)^{\!\!2}\right]
\sum_{n=1}^{\infty}(-1)^{n-1}e^{-\frac{n\hbar\omega}{2k_BT}}\right.
\nonumber \\
&&~~~
\left.
+\frac{1}{4}\left[1-\left(\frac{\Delta}{\hbar\omega}\right)^{\!\!2}\right]\!
\left[3+\left(\frac{\Delta}{\hbar\omega}\right)^{\!\!2}\right]\!
\left(\frac{v_Fk}{\omega}\right)^{\!\!2}\!
\left(\frac{\hbar\omega}{2k_BT}\right)^{\!\!2}
\sum_{n=1}^{\infty}(-1)^{n-1}n^2e^{-\frac{n\hbar\omega}{2k_BT}}\right\}.
\nonumber
\end{eqnarray}

Performing the summation according to Ref.~\cite{63}
\begin{eqnarray}
&&
\sum_{n=1}^{\infty}(-1)^{n-1}e^{-\frac{n\hbar\omega}{2k_BT}}
=\frac{1}{e^{\frac{\hbar\omega}{2k_BT}}+1},
\label{eq37} \\
&&
\sum_{n=1}^{\infty}(-1)^{n-1}n^2e^{-\frac{n\hbar\omega}{2k_BT}}
=\frac{e^{\frac{\hbar\omega}{2k_BT}}(e^{\frac{\hbar\omega}{2k_BT}}
-1)}{(e^{\frac{\hbar\omega}{2k_BT}}+1)^3},
\nonumber
\end{eqnarray}
\noindent
after identical transformations Eq.~(\ref{eq36}) can be written
in the following form:
\begin{eqnarray}
&&
{\rm Re}\Delta_T\sigma_{\|}(\omega,k,T)=-
\frac{2\sigma_0\theta(\hbar\omega\eta-\Delta)}{
e^{\frac{\hbar\omega}{2k_BT}}+1}\left\{
\vphantom{\left[\frac{{\rm tanh}\frac{\hbar\omega}{4k_BT}}{1+
e^{-\frac{\hbar\omega}{2k_BT}}}\right]}
1+\left(\frac{\Delta}{\hbar\omega}\right)^{\!\!2}+\frac{1}{2}
\left(\frac{v_Fk}{\omega}\right)^{\!\!2}\right.
\nonumber \\
&&~~~\left.
\times\left[1+3\left(\frac{\Delta}{\hbar\omega}\right)^{\!\!2}\right]
\!\left[1+
\frac{(\hbar\omega)^2-\Delta^2}{16(k_BT)^2}
\frac{{\rm tanh}\frac{\hbar\omega}{4k_BT}}{1+
e^{-\frac{\hbar\omega}{2k_BT}}}\right]\right\},
\nonumber \\[-2mm]
&&
\label{eq38}\\
&&
{\rm Re}\Delta_T\sigma_{\bot}(\omega,k,T)=-
\frac{2\sigma_0\theta(\hbar\omega\eta-\Delta)}{
e^{\frac{\hbar\omega}{2k_BT}}+1}\left\{
\vphantom{\left[\frac{{\rm tanh}\frac{\hbar\omega}{4k_BT}}{1+
e^{-\frac{\hbar\omega}{2k_BT}}}\right]}
1+\left(\frac{\Delta}{\hbar\omega}\right)^{\!\!2}-\frac{1}{2}
\left(\frac{v_Fk}{\omega}\right)^{\!\!2}\right.
\nonumber \\
&&~~~\left.
\times\left[1-\left(\frac{\Delta}{\hbar\omega}\right)^{\!\!2}\right]
\!\left[1-
\frac{3(\hbar\omega)^2+\Delta^2}{16(k_BT)^2}
\frac{{\rm tanh}\frac{\hbar\omega}{4k_BT}}{1+
e^{-\frac{\hbar\omega}{2k_BT}}}\right]\right\}.
\nonumber
\end{eqnarray}

As is seen from Eq.~(\ref{eq38}), the real parts of thermal
corrections
to the conductivity of gapped graphene are mostly determined
by the local contributions
\begin{equation}
{\rm Re}\Delta_T\sigma_{\|}(\omega,0,T)=
{\rm Re}\Delta_T\sigma_{\bot}(\omega,0,T)=-
\frac{2\sigma_0}{e^{\frac{\hbar\omega}{2k_BT}}+1}
\frac{(\hbar\omega)^2+\Delta^2}{(\hbar\omega)^2}
\theta(\hbar\omega-\Delta).
\label{eq39}
\end{equation}
\noindent
The nonlocal contributions in Eq.~(\ref{eq38}) become dominant
only at very low temperature when the complete thermal correction
is exponentially small.

By adding Eqs.~(\ref{eq21}) and (\ref{eq39}), one arrives at the
final expression for the real part of conductivity of gapped
graphene at nonzero temperature in the local approximation
\begin{equation}
{\rm Re}\sigma_{\|(\bot)}(\omega,0,T)=
\sigma_0
{\rm tanh}\frac{\hbar\omega}{4k_BT}
\frac{(\hbar\omega)^2+\Delta^2}{(\hbar\omega)^2}
\theta(\hbar\omega\eta-\Delta).
\label{eq40}
\end{equation}

\begin{figure}[b]
\vspace*{-1cm}
\centerline{\hspace*{2.5cm}
\includegraphics{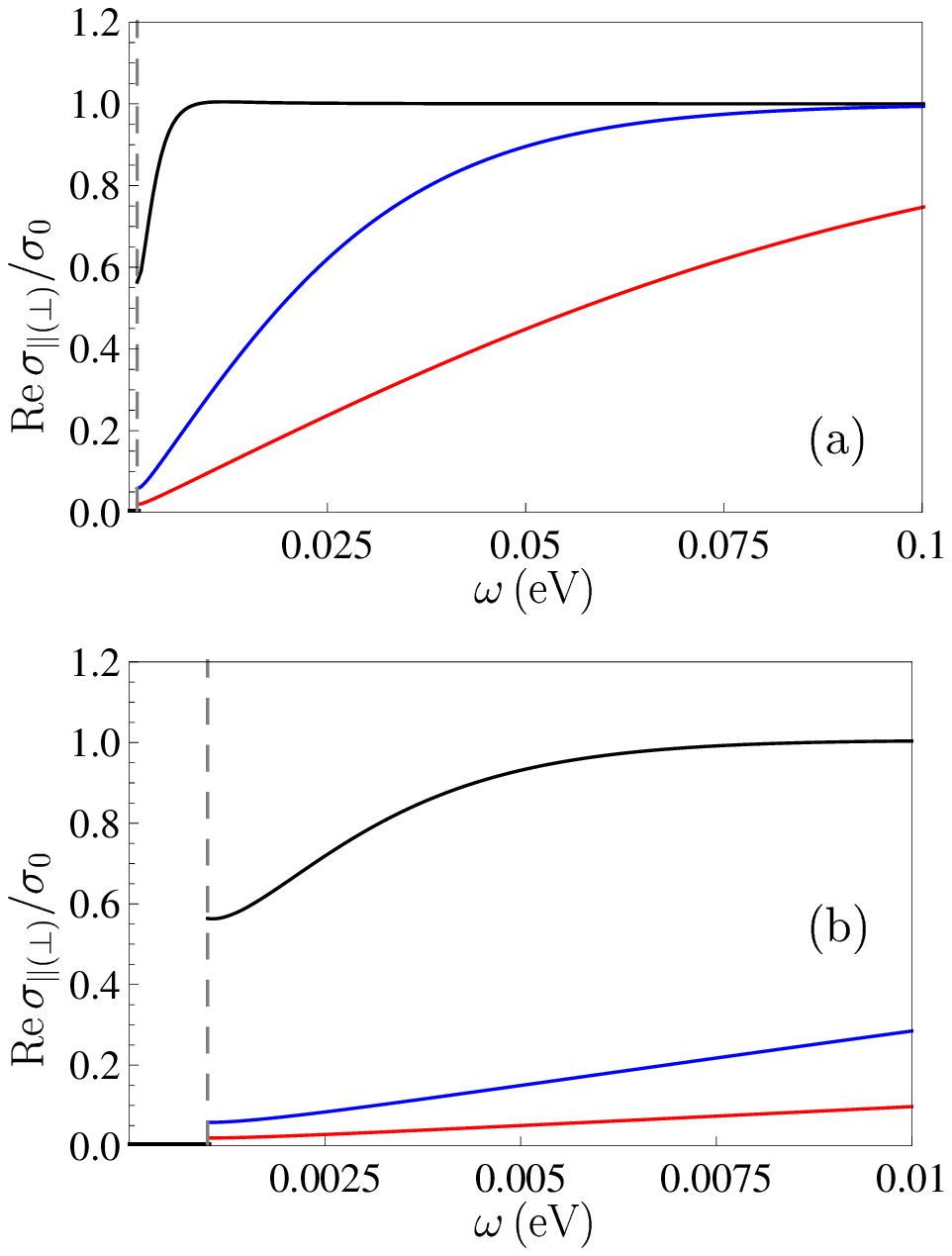}
}
\vspace*{-13cm}
\caption{\label{fg3}
(a) The normalized to $\sigma_0$ real
part of the conductivity of graphene with the gap parameter
$\Delta=0.001\,$eV is shown as a function of  frequency
by the three lines plotted from top to bottom at temperature
$T=10$, 100, and 300\,K, respectively.
The dashed line illustrates the condition $\hbar\omega=\Delta$.
(b) The same is shown on an enlarged scale at low frequencies.
}
\end{figure}

In Fig.~\ref{fg3}(a) we plot the real part of the normalized to
$\sigma_0$ longitudinal and transverse conductivities of
graphene with the gap parameter $\Delta=0.001\,$eV as a function of
frequency (measured in eV)
at different temperatures. The lines from top to
bottom correspond to $T=10$, 100, and 300\,K, respectively
(note that $T=300\,$K corresponds to $k_BT\approx 0.025\,$eV).
To gain a better understanding, the region of very low
frequencies is shown in Fig.~\ref{fg3}(b) on an enlarged scale.
According to Fig.~\ref{fg3}(a,b), the real part of conductivity
decreases with increasing $T$. This is because the thermal
correction to conductivity presented in Eq.~(\ref{eq39}) is
negative. As is seen in Fig.~\ref{fg3}(a,b), the real part of the
conductivity of graphene increases with increasing frequency
and goes to $\sigma_0$. This is, however, not a universal
property, but is determined by chosen (small) value of the gap
parameter.

\begin{figure}[b]
\vspace*{-1cm}
\centerline{\hspace*{2.5cm}
\includegraphics{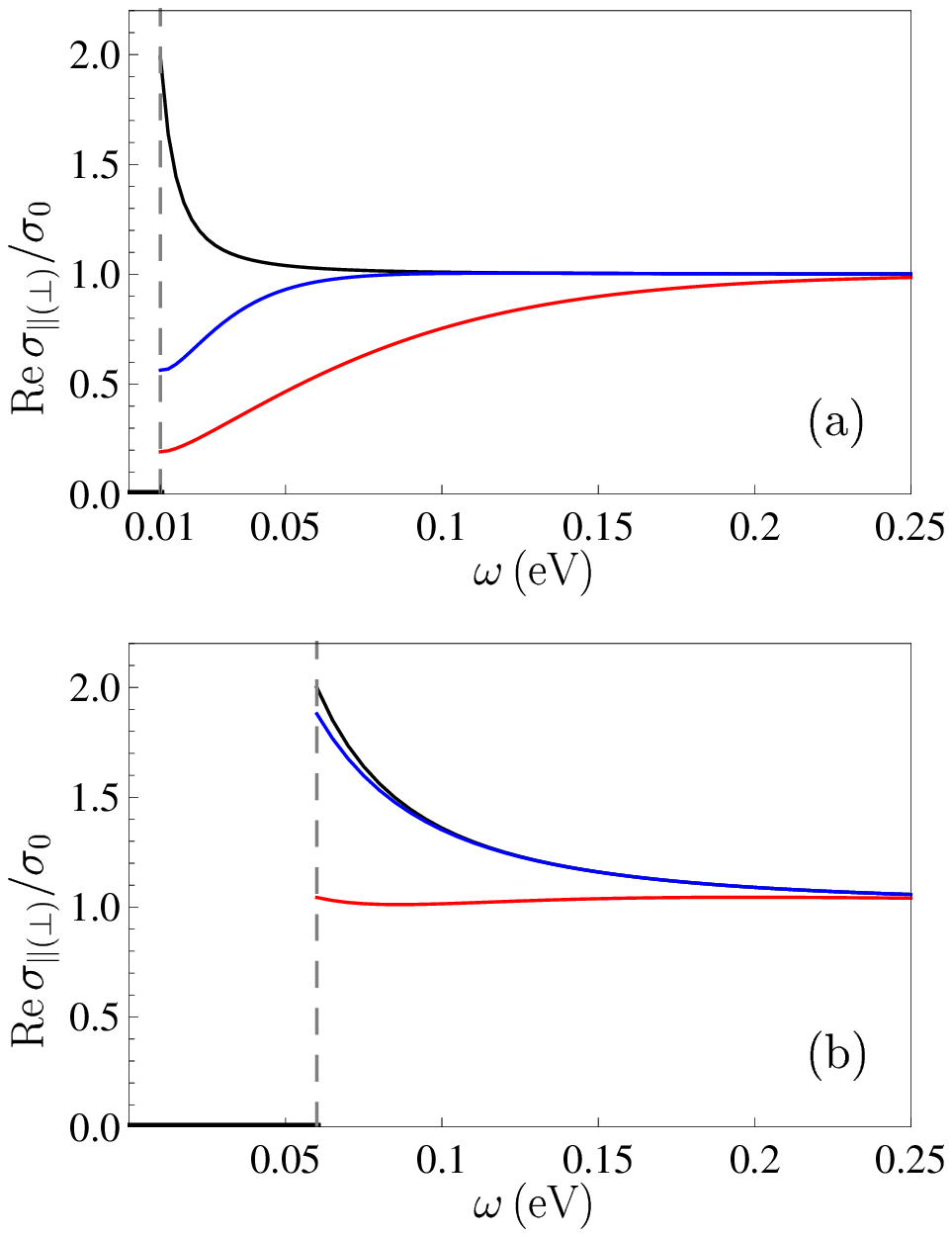}
}
\vspace*{-13cm}
\caption{\label{fg4}
The normalized to $\sigma_0$ real
part of the conductivity of graphene with the gap parameter
(a) $\Delta=0.01\,$eV and (b) $\Delta=0.06\,$eV
is shown as a function of  frequency
by the three lines plotted from top to bottom at temperature
$T=10$, 100, and 300\,K, respectively.
The dashed line illustrates the condition $\hbar\omega=\Delta$.
}
\end{figure}

To illustrate the latter statement, In Fig.~\ref{fg4} we plot
the real parts of the conductivities of graphene with the larger
gap parameters equal to (a) $\Delta=0.01\,$eV and (b)
$\Delta=0.06\,$eV as the functions of frequency. The lines from top
to bottom are plotted for $T=10$, 100, and 300\,K, respectively.
As is seen in Fig.~\ref{fg4}(a), at $T=300$ and 100\,K the real
part of conductivity increases, but at $T=10\,$K it decreases with
increasing frequency. Figure~\ref{fg4}(b) illustrates what is
happening with an increase of the gap parameter to
$\Delta=0.06\,$eV. In this case the real part of conductivity
at $T=10$ and 100\,K decreases with increasing frequency,
whereas at $T=300\,$K it becomes nonmonotonous due to an interplay
of two factors in Eq.~(\ref{eq40}).
With further increase of the gap parameter (up to
$\Delta=0.1\,$eV or larger) the real part of conductivity
becomes monotonously decreasing at all considered temperatures,
whereas the values of ${\rm Re}\sigma_{\|(\bot)}$ at $T=10$ and
100\,K become indistinguishable.

\section{Imaginary part of conductivity of gapped graphene at
nonzero temperature}

The imaginary parts of thermal corrections to the conductivity
of graphene are obtained from Eqs.~(\ref{eq2}) and (\ref{eq14})
\begin{eqnarray}
&&
{\rm Im}\Delta_T\sigma_{\|}(\omega,k,T)=-\frac{\omega}{4\pi\hbar k^2}
{\rm Re}\Delta_T\Pi_{00}(\omega,k,T),
\label{eq41} \\
&&
{\rm Im}\Delta_T\sigma_{\bot}(\omega,k,T)=\frac{c^2}{4\pi\hbar k^2\omega}
{\rm Re}\Delta_T\Pi(\omega,k,T).
\nonumber
\end{eqnarray}

At first we consider the frequency region $\hbar\omega\eta<\Delta$.
 Here, the real parts of the temperature corrections to the
polarization tensor are presented by Eq.~(\ref{eq9}). Thus, the
exact expressions for ${\rm Im}\Delta_T\sigma_{\|}$ and
${\rm Im}\Delta_T\sigma_{\bot}$ in this region are given by
Eqs.~(\ref{eq9}) and (\ref{eq41}). The total imaginary part of
the conductivity is given by
\begin{equation}
{\rm Im}\sigma_{\|(\bot)}(\omega,k,T)=
{\rm Im}\sigma_{\|(\bot)}^{(0)}(\omega,k)+
{\rm Im}\Delta_T\sigma_{\|(\bot)}(\omega,k,T),
\label{eq42}
\end{equation}
\noindent where ${\rm Im}\sigma_{\|(\bot)}^{(0)}$ is defined in
Eq.~(\ref{eq27}).

We expand Eq.~(\ref{eq9}) up to the first order in a small
parameter (\ref{eq19}) and find
\begin{eqnarray}
&&
{\rm Re}\Delta_T\Pi_{00}(\omega,k,T)=-
\frac{8\alpha\hbar c^2k^2}{\omega^2}
\int_{\Delta/(2\hbar c)}^{\infty}\frac{du}{e^{\beta u}+1}
\left[1+\frac{\Delta^2+(\hbar\omega)^2}{4(\hbar cu)^2-
(\hbar\omega)^2}\right],
\nonumber \\[-2mm]
&&
\label{eq43} \\
&&
{\rm Re}\Delta_T\Pi(\omega,k,T)=8\alpha\hbar k^2
\int_{\Delta/(2\hbar c)}^{\infty}\frac{du}{e^{\beta u}+1}
\left[1+\frac{\Delta^2+(\hbar\omega)^2}{4(\hbar cu)^2-
(\hbar\omega)^2}\right].
\nonumber
\end{eqnarray}

Substituting these equations in Eq.~(\ref{eq41}), one arrives at
equal results for ${\rm Im}\Delta_T\sigma_{\|}$ and
${\rm Im}\Delta_T\sigma_{\bot}$ in the local approximation
\begin{equation}
{\rm Im}\Delta_T\sigma_{\|(\bot)}(\omega,0,T)=\frac{8\sigma_0 c}{\pi\omega}
\int_{\Delta/(2\hbar c)}^{\infty}\frac{du}{e^{\beta u}+1}
\left[1+\frac{\Delta^2+(\hbar\omega)^2}{4(\hbar cu)^2-
(\hbar\omega)^2}\right].
\label{eq44}
\end{equation}
\noindent
Similar to Sec.~IV, it can be shown that the nonlocal
corrections to
${\rm Im}\Delta_T\sigma_{\|(\bot)}(\omega,0,T)$ are
negligibly small.

The total imaginary part of the conductivity of graphene in
the local approximation is given by
\begin{equation}
{\rm Im}\sigma_{\|(\bot)}(\omega,0,T)=
{\rm Im}\sigma_{\|(\bot)}^{(0)}(\omega,0)+
{\rm Im}\Delta_T\sigma_{\|(\bot)}(\omega,0,T),
\label{eq45}
\end{equation}
\noindent
where ${\rm Im}\sigma_{\|(\bot)}^{(0)}$ and
${\rm Im}\Delta_T\sigma_{\|(\bot)}$ are given by
Eqs.~(\ref{eq28}) and (\ref{eq44}), respectively.

Note that under the condition $\Delta\ll k_BT$ we can
obtain simple asymptotic expression for
${\rm Im}\Delta_T\sigma_{\|(\bot)}$ in Eq.~(\ref{eq44}).
For this purpose we take into account that the main contribution
to the integral is given by $u\approx1/\beta=k_BT/(\hbar c)$
and that the condition $\Delta\ll k_BT$ in the frequency region
under consideration leads to $\hbar\omega<\Delta\ll k_BT$ as
well. Then, one can neglect by the second term in square brackets
of Eq.~(\ref{eq44}) as compared to unity and, after the
integration, arrive at
\begin{equation}
{\rm Im}\Delta_T\sigma_{\|(\bot)}(\omega,0,T)\approx
8\sigma_0\frac{k_BT}{\pi\hbar\omega}\ln\left(1+
e^{-\frac{\Delta}{2k_BT}}\right)
\approx8\sigma_0\frac{k_BT}{\pi\hbar\omega}\ln 2.
\label{eq46}
\end{equation}

As is seen from Eqs.~(\ref{eq28}) and (\ref{eq44}), the imaginary
part of the conductivity of gapped graphene is negative, whereas
the thermal correction to it is positive. This means that the
total imaginary part of the conductivity may change its sign at some
frequency. Numerical computations confirm this conclusion.

\begin{figure}[b]
\vspace*{-1cm}
\centerline{\hspace*{2.5cm}
\includegraphics{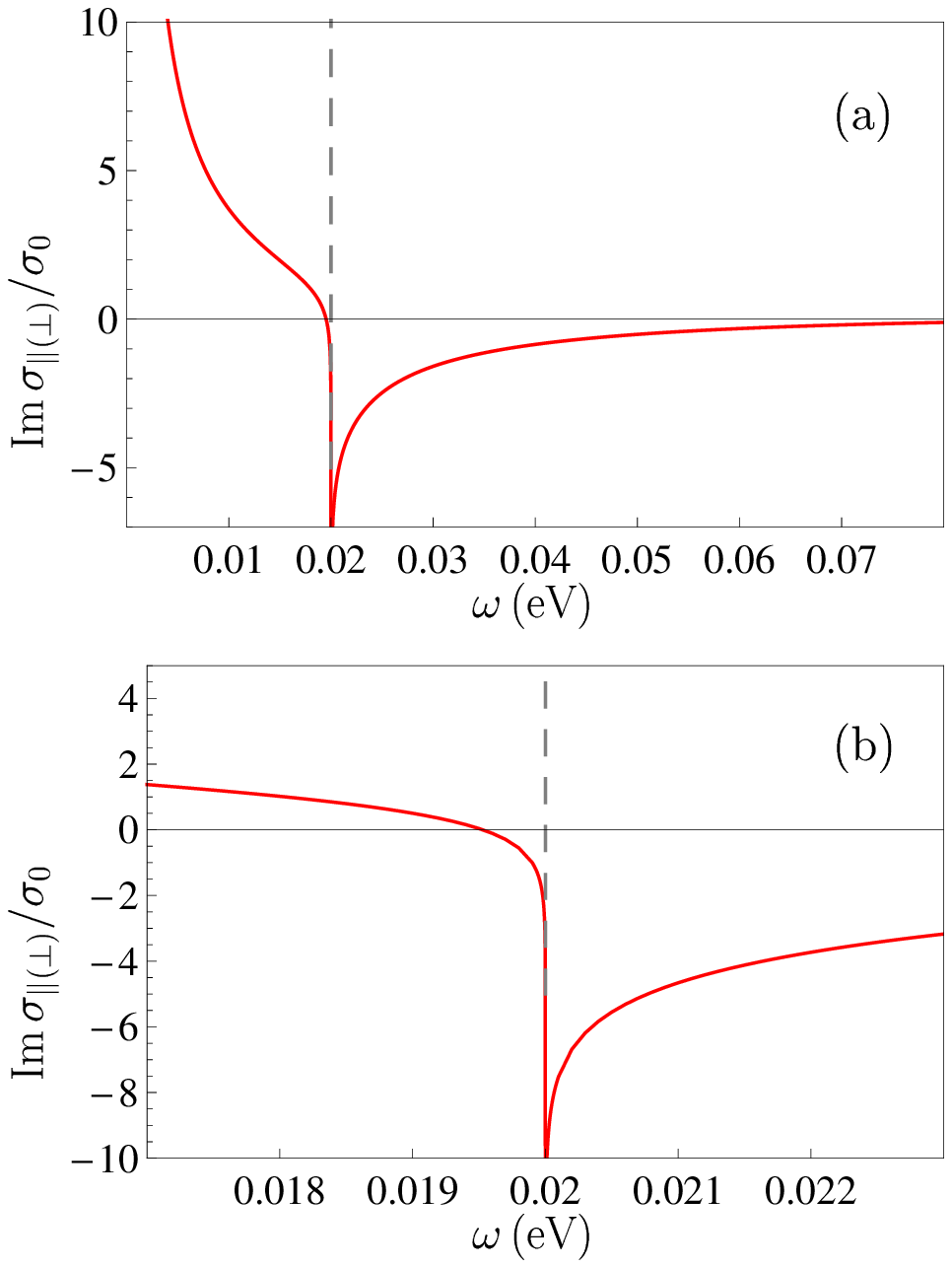}
}
\vspace*{-13cm}
\caption{\label{fg5}
(a) The normalized to $\sigma_0$ imaginary
part of the conductivity of graphene with the gap parameter
$\Delta=0.02\,$eV is shown as a function of  frequency
by the solid line plotted at temperature
$T=300\,$K.
The dashed line illustrates the condition $\hbar\omega=\Delta$.
(b) The same is shown on an enlarged scale in the vicinity
of $\hbar\omega=\Delta$.
}
\end{figure}

We have computed  ${\rm Im}\sigma_{\|(\bot)}$ normalized
to $\sigma_0$ as a function of $\omega$ in the region
$\hbar\omega<\Delta$ at room temperature $T=300\,$K in two ways:
by Eqs.~(\ref{eq9}), (\ref{eq27}), (\ref{eq41}), and (\ref{eq42}),
i.e., using the exact polarization tensor with several selected
values of $k$, and by Eqs.~(\ref{eq28}), (\ref{eq44}), and (\ref{eq45}),
i.e., in the local approximation, with nearly coincident results.
In Fig.~\ref{fg5}(a) the normalized imaginary part of conductivity
of graphene with the gap parameter $\Delta=0.02\,$eV is shown as
a function of frequency in the region $\hbar\omega<\Delta$ (the
latter is marked by the vertical dashed line). As is seen in
Fig.~\ref{fg5}(a),  ${\rm Im}\sigma_{\|(\bot)}$ varies
from infinity at zero frequency to minus infinity when
$\hbar\omega\to\Delta$. It has a root at some $\hbar\omega$
close to $\Delta$. For better vizualization, in Fig.~\ref{fg5}(b)
we show an immediate vicinity of the frequency
$\omega=\Delta/\hbar$ on an
enlarged scale. The root of ${\rm Im}\sigma_{\|(\bot)}$
is $\hbar\omega_0=0.019525\,$eV. The behavior of
${\rm Im}\Delta_T\sigma_{\|(\bot)}$ at $\omega=0$ is determined by
the thermal correction ${\rm Im}\Delta_T\sigma_{\|(\bot)}$,
whereas the behavior in the limiting case $\hbar\omega\to\Delta$
is caused
by the imaginary part of conductivity at zero temperature
${\rm Im}\sigma_{\|(\bot)}^{(0)}$ (compare with Fig.~\ref{fg2}).

\begin{figure}[b]
\vspace*{-6cm}
\centerline{\hspace*{.5cm}
\includegraphics{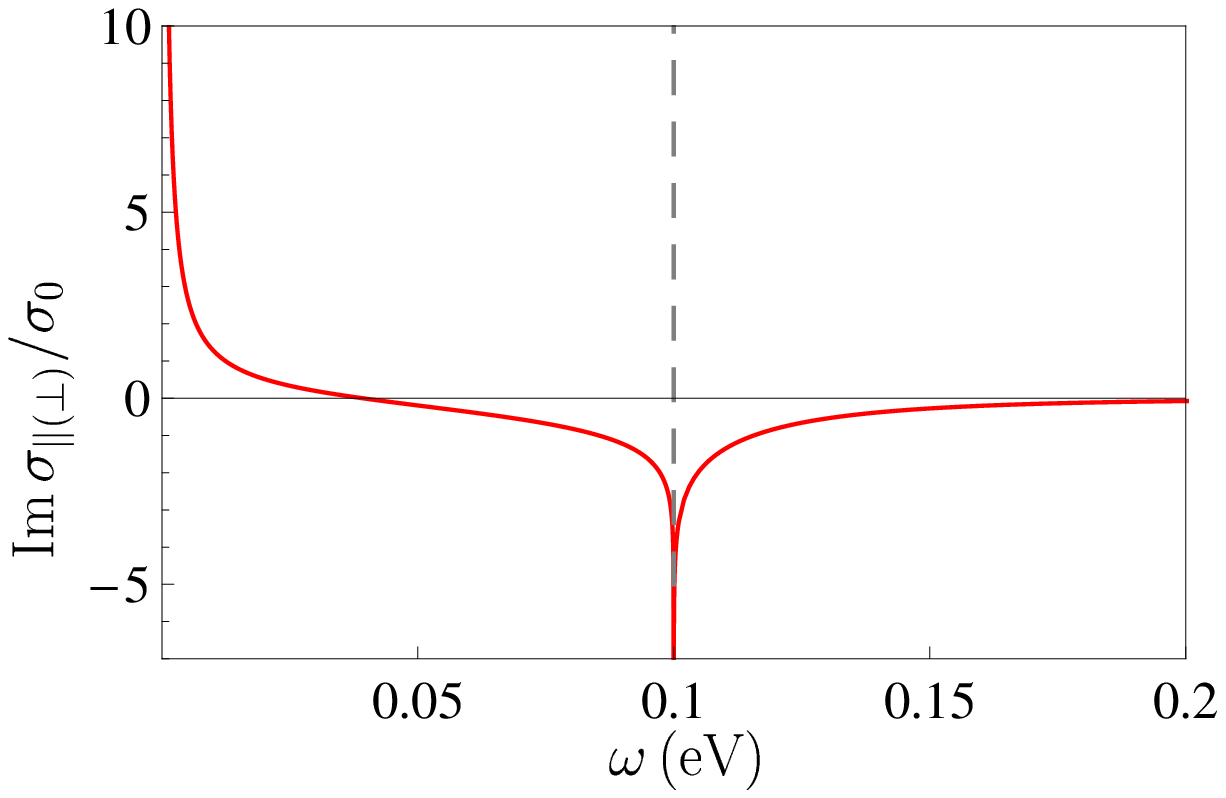}
}
\vspace*{-9cm}
\caption{\label{fg6}
The normalized to $\sigma_0$ imaginary
part of the conductivity of graphene with the gap parameter
$\Delta=0.1\,$eV is shown as a function of  frequency
by the solid line plotted at temperature
$T=300\,$K.
The dashed line illustrates the condition $\hbar\omega=\Delta$.
}
\end{figure}

For comparison purposes, in Fig.~\ref{fg6} the computational
results for
${\rm Im}\sigma_{\|(\bot)}$ at $T=300\,$K,
$\hbar\omega<\Delta$ are shown also for a larger gap
$\Delta=0.1\,$eV. As is seen in Fig.~\ref{fg6}, in this case  the
root of ${\rm Im}\sigma_{\|(\bot)}$, $\hbar\omega_0=0.03923\,$eV,
is much more separated from the border frequency
$\hbar\omega=\Delta$ than in Fig.~\ref{fg5} (see below for
the discussion of the frequency region $\hbar\omega\geq\Delta$).

We are coming now to the imaginary parts of the conductivities of
graphene ${\rm Im}\sigma_{\|(\bot)}$ under the condition
$\hbar\omega\geq\Delta$. The imaginary parts of the thermal
corrections, ${\rm Im}\Delta_T\sigma_{\|(\bot)}$, are again given
by Eq.~(\ref{eq41}) where, however, the quantities
${\rm Re}\Delta_T\Pi_{00}$ and ${\rm Re}\Delta_T\Pi$ are
expressed by Eqs.~(\ref{eq10})--(\ref{eq13}). These expressions,
together with the contribution at zero temperature in
Eq.~(\ref{eq27}),
are used below in numerical computations.

It is possible also to obtain the asymptotic expressions valid
under certain additional conditions. Thus, according to
Ref.~\cite{56}, at relatively high frequencies satisfying the
condition $\hbar\omega\gg k_BT\gg\Delta$ one obtains
\begin{eqnarray}
&&
{\rm Re}\Delta_T\Pi_{00}(\omega,k,T)\approx 48\alpha k^2
\frac{(k_BT)^3}{(\hbar\omega)^2}\,\frac{c}{\omega^2}
\zeta(3),
\nonumber \\
&&
{\rm Re}\Delta_T\Pi(\omega,k,T)\approx -48\alpha\frac{k^2}{c}
\,\frac{(k_BT)^3}{(\hbar\omega)^2}\zeta(3),
\label{eq47}
\end{eqnarray}
\noindent
where $\zeta(z)$ is the Riemann zeta function.

Substituting these equations in Eq.~(\ref{eq41}), we have
\begin{equation}
{\rm Im}\Delta_T\sigma_{\|(\bot)}(\omega,0,T)\approx
-\frac{48\zeta(3)}{\pi}\sigma_0\left(
\frac{k_BT}{\hbar\omega}\right)^3.
\label{eq48}
\end{equation}

If $\hbar\omega\gg k_BT$ and $\Delta\gg k_BT$ the following
asymptotic expressions are valid \cite{56}
\begin{eqnarray}
&&
{\rm Re}\Delta_T\Pi_{00}(\omega,k,T)\approx 24\alpha
\frac{k_BT}{(\hbar\omega)^2}\,\frac{k^2c\Delta^2}{\omega^2}
e^{-\frac{\Delta}{2k_BT}},
\nonumber \\
&&
{\rm Re}\Delta_T\Pi(\omega,k,T)\approx -24\alpha
\frac{k_BT}{(\hbar\omega)^2}\,\frac{k^2\Delta^2}{c}
e^{-\frac{\Delta}{2k_BT}}.
\label{eq49}
\end{eqnarray}
\noindent
Then from Eq.~(\ref{eq41}) one arrives at
\begin{equation}
{\rm Im}\Delta_T\sigma_{\|(\bot)}(\omega,0,T)\approx
-\frac{24}{\pi}\sigma_0
\frac{k_BT}{(\hbar\omega)^3}\Delta^2
e^{-\frac{\Delta}{2k_BT}}.
\label{eq50}
\end{equation}

The complete imaginary part of the conductivity of
gapped graphene under respective conditions is obtained by
summing Eqs.~(\ref{eq48}) and (\ref{eq50}) with Eq.~(\ref{eq28}).

Now we present the results of numerical computations using the
exact formulas (\ref{eq10})--(\ref{eq13}), (\ref{eq27}), and
(\ref{eq41}). In Figs.~\ref{fg5}(a) and \ref{fg5}(b) the
computational results for ${\rm Im}\sigma_{\|(\bot)}$
normalized to $\sigma_0$ are presented in the frequency
region $\hbar\omega>\Delta=0.02\,$eV at room temperature
$T=300\,$K. In Fig.~\ref{fg6} similar results are shown for
graphene with the gap parameter $\Delta=0.1\,$eV.
As is seen in Figs.~\ref{fg5}(a,b) and \ref{fg6}, in the
frequency region $\hbar\omega>\Delta$ the imaginary part
of $\sigma_{\|(\bot)}$ varies
from minus infinity to zero
with increasing frequency, i.e., qualitatively in the
same way as at zero temperature. By comparing
Figs.~\ref{fg5}(a) and \ref{fg6} it is seen that for a
larger gap parameter  ${\rm Im}\sigma_{\|(\bot)}$ faster
approaches to zero with increasing frequency.
It can be seen that Fig.~\ref{fg6} in the frequency region
$\hbar\omega>\Delta$ is in a rather good agreement with
the asymptotic expression (\ref{eq50}) added to the
zero-temperature contribution (\ref{eq28}). This is in
accordance with the application conditions of Eq.~(\ref{eq50}).

\section{Conclusions and discussion}

In the foregoing, we have investigated the conductivity of
gapped graphene described by the Dirac model on the basis of
first principles of quantum electrodynamics.
In doing so, no input parameter has been used.
Both cases of
zero and nonzero temperature have been considered.
The longitudinal and transverse conductivities of graphene
were expressed via the components of the polarization
tensor in (2+1)-dimensional space-time valid over the
entire plane of complex frequency \cite{54}.
This is a microscopic description in the sense that the
polarization tensor eventually represents the response of
individual electronic excitations to the electromagnetic
field.

At zero temperature, we have confirmed the analytic
expressions (\ref{eq21}) and (\ref{eq28}) for both real
and imaginary parts of the conductivity of graphene obtained
earlier in the local approximation \cite{27}. We have also
generalized these expressions
 in Eqs.~(\ref{eq17}) and (\ref{eq27}) taking into account nonlocal
effects.

The exact expressions for the real parts of thermal corrections
to the real parts of conductivities of gapped graphene at zero
temperature are defined in Eq.~(\ref{eq35}). The small role of
nonlocal effects is demonstrated in Eq.~(\ref{eq38}).
Finally, simple analytic expression for the real part of the
total conductivity of gapped graphene at any temperature is
given by Eq.~(\ref{eq40}). According to our results, the real
part of conductivity is equal to zero for $\hbar\omega<\Delta$.
In the frequency region $\hbar\omega\geq\Delta$ the real part
of conductivity takes some value $\sigma^{\ast}$
belonging to the interval $(0,2\sigma_0)$ at $\hbar\omega=\Delta$
and goes to $\sigma_0$ with increasing frequency.
The value $\sigma^{\ast}$ depends on the values of the
temperature and gap parameter.

For the imaginary parts of thermal corrections to the imaginary
parts of conductivities of gapped graphene at zero temperature,
the exact expressions are obtained in Eqs.~(\ref{eq41}) and
(\ref{eq9}) in the frequency region $\hbar\omega\eta<\Delta$.
In the local approximation, the simple Eq.~(\ref{eq44}) is
obtained, as well as the asymptotic expression (\ref{eq46}).
The most important impact of temperature on the imaginary part
of conductivity
of gapped graphene is that it goes to infinity when the frequency
vanishes (at $T=0$ the imaginary part of conductivity goes to
zero with vanishing frequency).

In the frequency region $\hbar\omega\eta\geq\Delta$ the exact
expressions for the imaginary parts of thermal corrections are
given by Eqs.~(\ref{eq41}) and (\ref{eq10})--(\ref{eq13}).
The simple asymptotic expressions in this case are given in
Eqs.~(\ref{eq48}) and (\ref{eq50}). In all cases, the results of
numerical computations performed using the exact formulas are
in a very good agreement with that using the local approximation
and with the asymptotic expansions in their areas of application.
In the limiting case $\Delta\to 0$ all the results of this paper
smoothly transform into the previously obtained ones for pure
graphene \cite{58}.

Finally, we conclude that the description of graphene by means
of the polarization tensor calculated using the first principles
of quantum electrodynamics at nonzero temperature in
(2+1)-dimensional space-time turns out to be very fruitful
not only in the Casimir effect, but also for a better understanding
of more conventional physical phenomena, such as the reflectivity
and conductivity properties of graphene. In this respect it
would be interesting to apply the results of
the recent Ref.~\cite{64}
 for theoretical description of electrical conductivity of
 graphene with nonzero chemical potential using the formalism
 of the polarization tensor.

\section*{Acknowledgments}

The work of V.M.M.~was partially supported by the Russian Government
Program of Competitive Growth of Kazan Federal University.

%%%%%%%%%%%%%%%%%%%%%%%%%%%

%%%%%%%%%%%%%%%%%%%%%%%%%%

%\end{document}
%%%%%%%%%%%%%%%%%%%%%%%%%%%%%
\newpage

%%%%%%%__FIGURE__1__%%%%%%%%%%%%%%%%%%%%
\begin{figure}[b]
\vspace*{-6cm}
\centerline{\hspace*{2.5cm}
\includegraphics{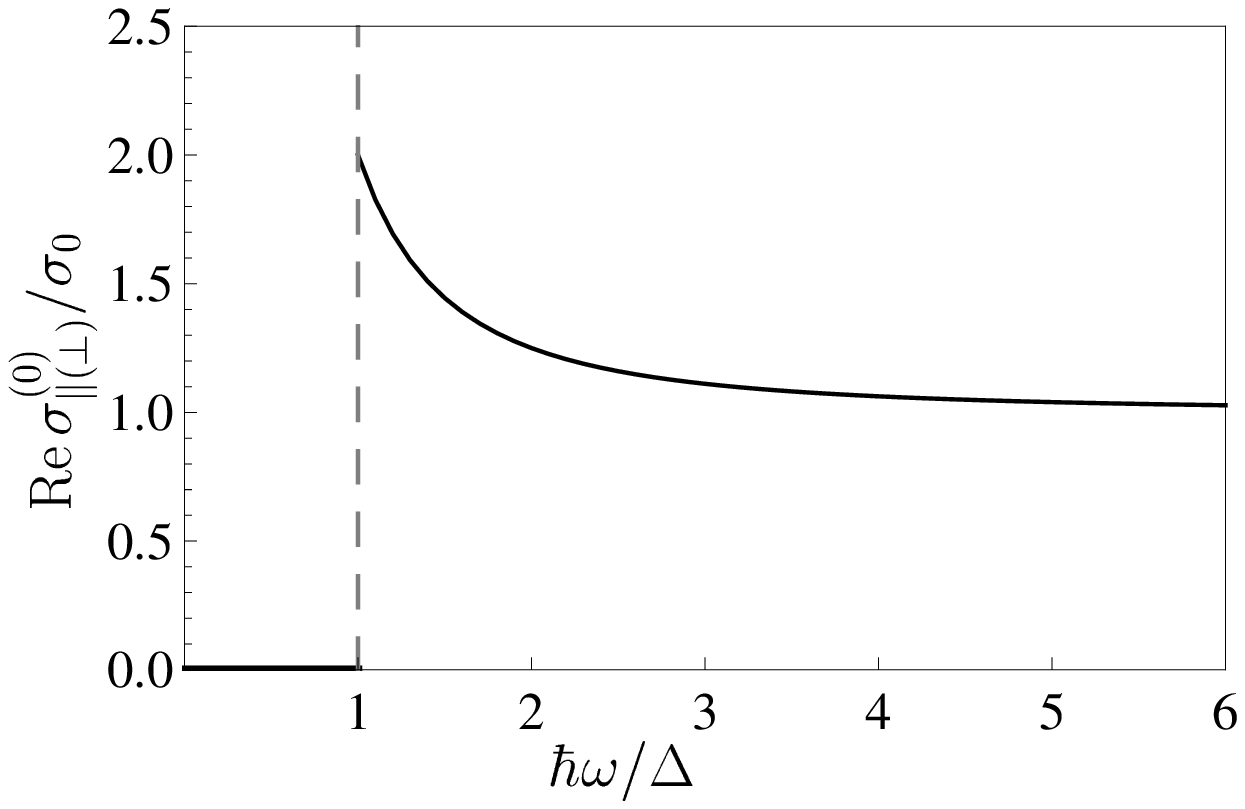}
}
\vspace*{-9cm}
\caption{\label{fg1}
The real part of the conductivity of gapped graphene
at zero temperature normalized to $\sigma_0$ is shown by the solid line
as a function of frequency divided by the width of the gap.
The dashed line illustrates the condition $\hbar\omega=\Delta$.
}
\end{figure}
%%%%%%%%%%%%%
%%%%%%%__FIGURE__2__%%%%%%%%%%%%%%%%%%%%
%%%%%%%%%%%%%
%%%%%%%__FIGURE__3__%%%%%%%%%%%%%%%%%%%%
%%%%%%%%%%%%%
%%%%%%%__FIGURE__4__%%%%%%%%%%%%%%%%%%%%
%%%%%%%%%%%%%
%%%%%%%__FIGURE__5__%%%%%%%%%%%%%%%%%%%%
%%%%%%%%%%%%%
%%%%%%%__FIGURE__6__%%%%%%%%%%%%%%%%%%%%
%%%%%%%%%%%%%

\begin{thebibliography}{99}
\bibitem{1}
M.~I.~Katsnelson, {\it Graphene: Carbon in Two Dimensions} (Cambridge University
Press, Cambridge, 2012).
\bibitem{2}
A.~H.~Castro Neto, F.~Guinea, N.~M.~R.~Peres, K.~S.~Novoselov, and A.~K.~Geim,
Rev. Mod. Phys. {\bf 81}, 109 (2009).
\bibitem{3}
N.~M.~R.~Peres, Rev. Mod. Phys. {\bf 82}, 2673 (2010).
\bibitem{4}
S.~Das~Sarma, S.~Adam, E.~H.~Hwang, and E.\ Rossi,
Rev. Mod. Phys. {\bf 83}, 407 (2011).
\bibitem{5}
A.~W.~W.~Ludwig, M.~P.~A.~Fisher, R.~Shankar, and G.\ Grinstein,
Phys. Rev. B {\bf 50}, 7526 (1994).
\bibitem{6}
T.~Ando, Y.~Zheng, and H.~Suzuura,
J. Phys. Soc. Jpn. {\bf 71}, 1318 (2002).
\bibitem{7}
V.~P.~Gusynin and S.~G.~Sharapov, Phys. Rev. B {\bf 73},
245411 (2006).
\bibitem{8}
M.~I.~Katsnelson, Eur. Phys. J. B {\bf 51}, 157 (2006).
\bibitem{9}
K.~Ziegler, Phys. Rev. Lett. {\bf 97}, 266802 (2006).
\bibitem{10}
K.~Ziegler, Phys. Rev. B {\bf 75}, 233407 (2007).
\bibitem{11}
V.~P.~Gusynin, S.~G.~Sharapov, and J.~P.~Carbotte, Phys. Rev. Lett. {\bf 98},
157402 (2007).
\bibitem{12}
M.~Trushin and J.~Schliemann,
Phys. Rev. Lett. {\bf 99}, 216602 (2007).
\bibitem{13}
L.~A.~Falkovsky and A.~A.~Varlamov, Eur. Phys. J. B {\bf 56}, 281 (2007).
\bibitem{14}
L.~A.~Falkovsky and S.~S.~Pershoguba,
Phys. Rev. B {\bf 76}, 153410 (2007).
\bibitem{15}
M.~Auslender and M.~I.~Katsnelson,
Phys. Rev. B {\bf 76}, 235425 (2007).
\bibitem{16}
V.~P.~Gusynin and S.~G.~Sharapov,
J. Phys.: Condens. Matter {\bf 19}, 026222 (2007).
\bibitem{17}
V.~P.~Gusynin, S.~G.~Sharapov, and J.~P.~Carbotte,
Int. J. Mod. Phys. B {\bf 21}, 4611 (2007).
\bibitem{18}
T.~Stauber, N.~M.~R.~Peres, and A.~K.~Geim, Phys. Rev. B {\bf 78}, 085432 (2008).
\bibitem{19}
N.~M.~R.~Peres and T.~Stauber,
Int. J. Mod. Phys. B {\bf 22}, 2529 (2008).
\bibitem{20}
T.~G.~Pedersen, A.-P.~Jauho, and K.~Pedersen,
Phys. Rev. B {\bf 79}, 113406 (2009).
\bibitem{21}
M.~Lewkowicz and B.~Rosenstein,
Phys. Rev. Lett. {\bf 102}, 106802 (2009).
\bibitem{22}
J.~J.~Palacios,
Phys. Rev. B {\bf 82}, 165439 (2010).
\bibitem{23}
L.~Moriconi and D.~Niemeyer,
Phys. Rev. B {\bf 84}, 193401 (2011).
\bibitem{24}
P.~V.~Buividovich, E.~V.~Luschevskaya, O.~V.~Pavlovsky, M.\ I.\ Polikarpov,
and M.\ V.\ Ulybyshev,
Phys. Rev. B {\bf 86}, 045107 (2012).
\bibitem{25}
{\' A}.~B{\' a}csi and A.~Virosztek,
Phys. Rev. B {\bf 87}, 125425 (2013).
\bibitem{26}
C.~A.~Dartora and G.~G.~Cabrera,
Phys. Rev. B {\bf 87}, 165416 (2013).
\bibitem{27}
T.~Stauber,
J. Phys.: Condens. Matter {\bf 26}, 123201 (2014).
\bibitem{28}
T.~Louvet, P.~Delplace, A.~A.~Fedorenko, and D.\ Carpentier,
Phys. Rev. B {\bf 92}, 155116 (2015).
\bibitem{29}
D.~K.~Patel, A.~C.~Sharma, and S.\ S.\ Z.\ Ashraf,
Phys. Status Solidi {\bf 252}, 282 (2015).
\bibitem{30}
M.~Merano,
Phys. Rev. A {\bf 93}, 013832 (2016).
\bibitem{31}
Y.-W.~Tan, Y.~Zhang, K.~Bolotin, Y.~Zhao, S.~Adam, E.~H.~Hwang,
S.~Das~Sarma, H.\ L.\ Stormer, and P.\ Kim,
Phys. Rev. Lett. {\bf 99}, 246803 (2007).
\bibitem{32}
R.~R.~Nair, P.~Blake, A.~N.~Grigorenko, K.\ S.\ Novoselov, T.\ J.\ Booth,
T.~Stauber, N.~M.~R.~Peres, and A.~K.~Geim,
Science {\bf 320}, 1308 (2008).
\bibitem{33}
Z.~Li, E.~Henriksen, Z.~Jiang, Z.~Hao, M.\ Martin, P.\ Kim, H.\ Stormer,
and D.~Basov,
Nature Phys. {\bf 4}, 532 (2008).
\bibitem{34}
K.~F.~Mak, M.~Y.~Sfeir, Y.~Wu, C.\ H.\ Lui, J.\ A.\ Misewich,
and T.\ F.\ Heinz,
Phys. Rev. Lett. {\bf 101}, 196405 (2008).
\bibitem{35}
J.~Horng, C.-F.~Chen, B.~Geng {\it et al.},
Phys. Rev. B {\bf 83}, 165113 (2011).
\bibitem{36}
M.~Bordag, I.~V.~Fialkovsky, D.~M.~Gitman, and D.~V.~Vassilevich, Phys. Rev. B
{\bf 80}, 245406 (2009).
\bibitem{37}
I.~V.~Fialkovsky, V.~N.~Marachevsky, and D.~V.~Vassilevich, Phys. Rev. B {\bf 84},
035446 (2011).
\bibitem{37a}
G.~L.~Klimchitskaya, U.~Mohideen, and V.~M.~Mostepanenko, Rev. Mod. Phys.
{\bf 81}, 1827 (2009).
\bibitem{37b}
M.~Bordag, G.~L.~Klimchitskaya, U.~Mohideen, and V.~M.~Mostepanenko,
{\it Advances in the Casimir Effect} (Oxford University Press, Oxford, 2015).
\bibitem{38}
M.~Bordag, G.~L.~Klimchitskaya, and V.~M.~Mostepanenko, Phys. Rev. B {\bf 86},
165429 (2012).
\bibitem{39}
M.~Chaichian, G.~L.~Klimchitskaya, V.~M.~Mostepanenko, and A.~Tureanu,
Phys. Rev. A {\bf 86}, 012515 (2012).
\bibitem{40}
G.~L.~Klimchitskaya and V.~M.~Mostepanenko, Phys. Rev. B {\bf 87}, 075439 (2013).
\bibitem{41}
G.~L.~Klimchitskaya, U.~Mohideen, and V.~M.~Mostepanenko, Phys. Rev B {\bf 89},
115419 (2014).
\bibitem{42}
G.~L.~Klimchitskaya, V.~M.~Mostepanenko, and Bo E.~Sernelius,
Phys. Rev. B {\bf 89}, 125407 (2014).
\bibitem{43}
B.~Arora, H.~Kaur, and B.~K.~Sahoo,
J. Phys. B {\bf 47}, 155002 (2014).
\bibitem{44}
K.~Kaur, J.~Kaur, B.~Arora,  and B.~K.~Sahoo,
Phys. Rev. B {\bf 90}, 245405 (2014).
\bibitem{45}
G.~L.~Klimchitskaya and V.~M.~Mostepanenko, Phys. Rev. B {\bf 91}, 045412 (2015).
\bibitem{46}
G.~G\'{o}mez-Santos, Phys. Rev. B {\bf 80}, 245424 (2009).
\bibitem{47}
D.~Drosdoff and L.~M.~Woods, Phys. Rev. B {\bf 82}, 155459 (2010).
\bibitem{48}
D.~Drosdoff and L.~M.~Woods, Phys. Rev. A {\bf 84}, 062501 (2011).
\bibitem{49}
Bo E.~Sernelius, Europhys. Lett. {\bf 95}, 57003 (2011).
\bibitem{50}
Bo E.~Sernelius, Phys. Rev. B {\bf 85}, 195427 (2012).
\bibitem{51}
A.~D.~Phan, L.~M.~Woods, D.~Drosdoff, I.~V.~Bondarev, and N.~A.~Viet, Appl.
Phys. Lett. {\bf 101}, 113118 (2012).
\bibitem{52}
G.~L.~Klimchitskaya and V.~M.~Mostepanenko,
Phys. Rev A {\bf 89}, 052512 (2014).
\bibitem{53}
A.~A.~Banishev, H.~Wen, J.~Xu, R.~K.~Kawakami,
G.~L.~Klimchitskaya, V.~M.~Mostepanenko, and U.~Mohideen,
Phys. Rev. B {\bf 87}, 205433 (2013).
\bibitem{54}
M.~Bordag, G.~L.~Klimchitskaya, V.~M.~Mostepanenko, and V.~M.~Petrov,
Phys. Rev. D {\bf 91}, 045037 (2015); {\bf 93}, 089907(E) (2016).
\bibitem{55}
G.~L.~Klimchitskaya and V.~M.~Mostepanenko,
Phys. Rev B {\bf 91}, 174501 (2015).
\bibitem{55a}
G.~L.~Klimchitskaya,
Int. J. Mod. Phys. A {\bf 31}, 1641026 (2016).
\bibitem{57}
G.~L.~Klimchitskaya, C.~C.~Korikov, and V.~M.~Petrov,
Phys. Rev. B {\bf 92}, 125419 (2015); {\bf 93}, 159906(E) (2016).
\bibitem{58}
G.~L.~Klimchitskaya and V.~M.~Mostepanenko,
Phys. Rev B {\bf 93}, 245419 (2016).
\bibitem{56}
G.~L.~Klimchitskaya and V.~M.~Mostepanenko,
Phys. Rev A {\bf 93}, 052106 (2016).
\bibitem{59}
P.~K.~Pyatkovsky, J. Phys.: Condens. Matter {\bf 21}, 025506 (2009).
\bibitem{60}
V.~P.~Gusynin, S.~G.~Sharapov, and J.~P.~Carbotte, New J. Phys. {\bf 11},
095013 (2009).
\bibitem{61}
S.~A.~Jafari, J. Phys.: Condens. Matter {\bf 24}, 205802 (2012).
\bibitem{62}
Bo E.~Sernelius,
J. Phys.: Condens. Matter {\bf 27}, 214017 (2015).
\bibitem{63}
I.~S.~Gradshtein and I.~M.~Ryzhik,
{\it Table of Integrals, Series and Products}
(Academic Press, New York, 1980).
\bibitem{64}
M.~Bordag, I.~Fialkovsky, and D.~Vassilevich,
Phys. Rev. B {\bf 93}, 075414 (2016).
\end{thebibliography}
\end{document}